# Mapping the local spatial charge in defective diamond by means of NV sensors – A "self-diagnostic" concept

J. Forneris[1,2], S. Ditalia Tchernij[2,1], P. Traina[3], E. Moreva[3], N. Skukan[4], M. Jakšić[4], V. Grilj[4], L. Croin[3], G. Amato[3], I.P. Degiovanni[3], B. Naydenov[5], F. Jelezko[5], M. Genovese[3,1], P. Olivero[2,1]

[1]*Istituto Nazionale di Fisica Nucleare (INFN), Sez. Torino, via P. Giuria 1, 10125, Torino, Italy*
[2]*Physics Department and "NIS " Inter-departmental Centre - University of Torino; via P. Giuria 1, 10125, Torino, Italy*
[3]*Istituto Nazionale di Ricerca Metrologica (INRiM); Strada delle Cacce 91, 10135 Torino, Italy*
[4]*Ruđer Bošković Institute, Bijenicka 54, P.O. Box 180, 10002 Zagreb, Croatia*
[5]*Institute for Quantum Optics and Center for Integrated Quantum Science and Technology (IQST), Albert-Einstein-Allee 11, Universität Ulm, D-89069 Ulm, Germany*

**ABSTRACT**

Electrically-active defects have a significant impact on the performance of electronic devices based on wide band-gap materials such as diamond. This issue is ubiquitous in diamond science and technology, since the presence of charge traps in the active regions of different classes of diamond-based devices (detectors, power diodes, transistors) can significantly affect their performances, due to the formation of space charge, memory effects and the degradation of the electronic response associated with radiation damage. Among the most common defects in diamond, the nitrogen-vacancy (NV) center possesses unique spin properties which enable high-sensitivity field sensing at the nanoscale. Here we demonstrate that NV ensembles can be successfully exploited to perform a direct local mapping of the internal electric field distribution of a graphite-diamond-graphite junction exhibiting electrical properties dominated by trap- and space-charge-related conduction mechanisms. By performing optically-detected magnetic resonance measurements, we performed both punctual readout and spatial mapping of the electric field in the active region at different bias voltages. In this novel "self-diagnostic" approach, defect complexes represent not only the source of detrimental space charge effects, but also a unique tool to directly investigate them, by providing experimental evidences on the conduction mechanisms that in previous studies could only be indirectly inferred on the basis of conventional electrical and optical characterization.

## I. INTRODUCTION

Diamond is an appealing material for the development of innovative devices, such as high-power and fast electronics [1-4], radiation dosimeters [5,6] and detectors [7-8], biosensors [9-10] and more recently integrated platforms for quantum technologies [10-14]. For all of the above applications, a major issue to be addressed in the optimization of the device performance is represented by the effect of lattice defects on the electrical properties of the material. It has been extensively reported that the introduction of trap levels caused by the interaction with energetic radiation causes electric field (EF) inhomogeneities and polarization/memory effects in the material [15-17]. In previous works these effects were widely investigated by the analysis of current-voltage characteristics exhibiting complex non-ohmic behaviors, elucidating several conduction models ranging from Space Charge Limited Current (SCLC) to Poole-Frenkel (PF) mechanisms [18-21]. Complementarily, Ion/Electron/X-ray/VIS Beam Induced Charge microscopies allow the mapping of the charge transport parameters of diamond at the microscale [17,22-24]. Although insightful, the above-mentioned techniques cannot provide a direct and unequivocal experimental evidence of the local EF distribution in the defective material, but rather need to assume it in simplified Finite Elements models. Among the large variety of lattice defects in diamond, the nitrogen-vacancy complex (NV center) emerged as a system characterized by unique spin properties, enabling magnetic, thermal and electric field sensing with high sensitivity and spatial resolution [25-30]. The very same NV centers which are created (among other types of defect complexes) by radiation damage [31] can be therefore exploited to locally investigate the internal EF distribution of the irradiated devices [30,32]. In this work, we show that the optically-detected magnetic resonance (ODMR) readout of the electronic spin from NV center ensembles can be usefully exploited to map the internal EF distribution in defective diamond, in order to diagnose the effects of electrically-active centers. To this purpose, we employed a graphite-diamond-graphite junction fabricated by means of MeV ion beam lithography [Carbon]. High ($\sim 10^5$ V cm$^{-1}$) local EFs were applied in the micrometer-sized diamond region comprised between the graphitic electrodes, containing a high (i.e. $\sim 10^{15}$ cm$^{-3}$) density of radiation induced defects. In this "self-diagnostic" approach, it was thus possible to investigate space charge buildup and the local EF distribution under different biasing conditions.

## II. EXPERIMENTAL

We investigated a graphite-diamond-graphite junction embedded in a CVD "optical grade" type IIa single-crystal substrate with <100> orientation (**Fig. 1a**). The structure consists of two ∼15×100 μm² graphitic micro-electrodes elongated in the <100> crystal direction, whose endpoints are spaced by a ∼9 μm diamond gap (an optical micrograph is shown in **Fig.1b**). The electrodes were fabricated by raster scanning a ∼5 μm sized 6 MeV $C^{3+}$ beam, as explained in details in previous works [21,33]. The implantation fluence (∼4×10¹⁶ cm⁻²) ensured the amorphization of a diamond layer at the end of ion range (~2.7 μm) [34], which was converted to a graphitic phase by a subsequent thermal treatment. The ion microbeam fabrication also resulted in the implantation of a "halo" of stray ions surrounding the graphitized layers, having a size comparable with the inter-electrode gap and associated with an estimated vacancy density of ∼1×10²² cm⁻³ [21,34]. Such a value is below the "graphitization threshold" for deep MeV ion implantation (6-9×10²² cm⁻³, [35,36]), but high enough to promote the formation of a high concentration of NV centers (i.e. 10³–10⁴ centers μm⁻³ [21]) upon thermal annealing [33], as well as to introduce other types of deep trap levels. The lattice structure was not fully recovered by the thermal treatment and the overall traps concentration in the same device was quantified as ~2×10¹⁵ cm⁻³ by previous electrical characterizations [33].

Photoluminescence (PL) and ODMR measurements were performed using a confocal microscope [14] with 515 nm laser excitation. The PL emission reported in the following maps and ODMR spectra was acquired in the 650 – 800 nm spectral range. Microwave excitation was sourced by a Cu wire (∅ ∼30 μm) placed on the sample surface at a distance of ∼25 μm from the active region and connected to an Agilent EXG Vector Signal Generator through a ZHL-16W-S+ amplifier from Minicircuits.

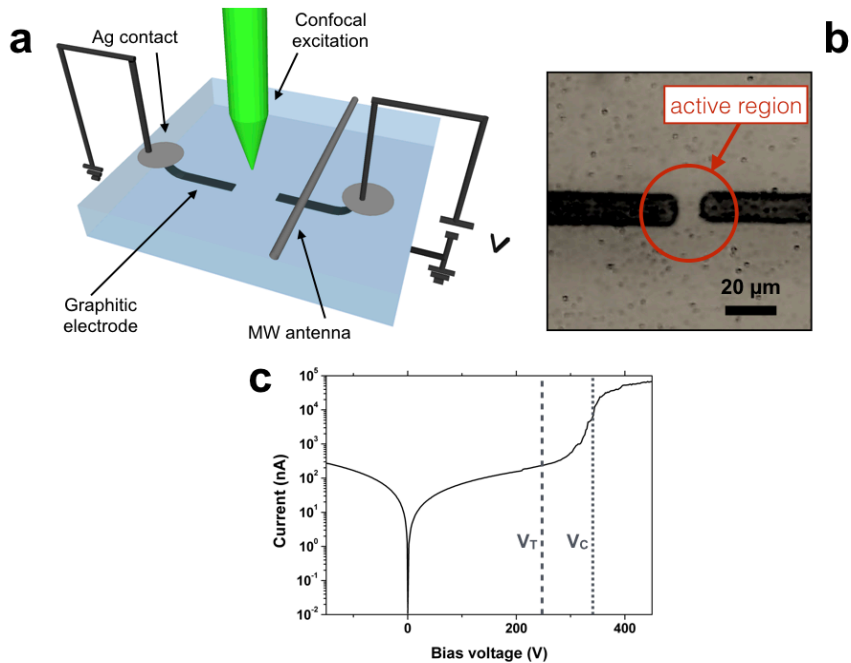

**Figure 1**: **a)** Schematic representation (not to scale) of the graphite-diamond-graphite junction. **b)** Optical micrograph of the device, in which the active inter-electrode region is highlihted. Current-voltage characteristic of the junction: **c)** in the -150 – +450V bias voltage range (semi-logarithmic scale).

### III. RESULTS AND DISCUSSION

*Electrical characterization*. The room-temperature current-voltage characteristic of the device is presented in **Figure 1c** in the same voltage ranges explored in the ODMR measurements [21,33]. The curve exhibits an ohmic trend up to a voltage threshold ($V_T$ ~ +250 V under positive bias), where the space charge density starts affecting the conduction mechanism and a deviation from the linear trend is observed. Consistently with the SCLC model [21,33], this deviation is associated with the progressive filling of empty traps by the electrons injected the inter-electrode gap [37], and leads to a transition to a high-current regime above a critical value $V_C$ (∼ +330 V). At high bias voltages, the super-linear dependence of the current on the applied bias is compatible with the PF conduction model [37].

According to the SCLC model, the current is limited at biases in the $V_T < V < V_C$ range by a counter-field built by the progressive filling of traps by carriers injected in the active region of the device [38]. Such an interpretation is based on a unipolar conduction mechanism and is consistent with several experimental features observed over repeated measurement cycles, i.e.: different slopes in the current-voltage curves under polarity reversal (**Fig. 1c**); a progressive conversion of the NV centers to a negative charge state at increasing currents [21]; a decrease in the critical voltage at increasing temperatures [33]; a marginal (i.e. few V) variability in the value of $V_C$ from different experimental runs [33]; a pronounced current-voltage hysteresis at decreasing bias voltages [14].

*ODMR spectra at variable applied bias voltage.* The PL mapping of the device at zero bias voltage (**Fig. 2a**) exhibited an intense PL emission from the regions located above the electrodes, associated with radiation-induced defects between the sample surface and the graphitic layer. Conversely, the PL emission from the active region is largely dominated by NV ensembles [21]. Before ODMR measurements, the junction underwent multiple voltage sweeps in the 0 – +200V range, in order to progressively build a space charge in the active region. Between **Figs. 2b-2d**, the label "[C]" identifies the initial "charged" state of the device at zero bias voltage. Subsequently, ODMR spectra were acquired at increasing bias voltages from 0 V to +350 V at the fixed position labeled by the green spot in **Fig. 2a**: this first voltage sweep is indicated between **Figs. 2b-2d** by the arrow labeled as "1". A second bias voltage sweep from 0 V to -105 V (labeled as "2" between **Figs. 2b-2d**) was subsequently acquired from the same spot. Each ODMR spectrum (**Fig. 2c** reports a typical one collected at $V_{bias}$ = +25 V) displays two dips at the characteristic resonance frequencies for NV$^-$ centers, with a splitting value which is systematically dependent on the intensity of the local EF, as clearly visible in **Fig. 2b**, where the whole set of spectra is encoded in color scale as a function of the applied bias voltage.

The data were interpreted according to the Hamiltonian of the ground state of the NV$^-$ system, which describes the energy shift of the $m_s = \pm 1$ electronic levels due to the spin (*S*) interaction with static magnetic (*B*), electric (*E*) and strain (*F*) fields [30,39]:

$$H = H_0 + \underbrace{g\mathbf{S}\cdot\mathbf{B}}_{\alpha} + \underbrace{d_{\parallel}(E_z + F_z)\left[S_z^2 + S(S+1)/3\right]}_{\beta} - \underbrace{d_{\perp}(E_x + F_x)(S_xS_y + S_yS_x)}_{\gamma} - \underbrace{d_{\perp}(E_y + F_y)(S_x^2 - S_y^2)}_{\delta} \quad (1)$$

where $H_0$ defines the fine structure of the NV$^-$ ground state. The term $\alpha$ describes the Zeeman splitting of the ODMR resonances, while the following three terms account for the Stark effect due to the coupling with EFs which are parallel ($\beta$) and perpendicular ($\gamma,\delta$) to the defect axis, with coupling constants $d_{\perp}$ = 17 Hz V$^{-1}$ cm and $d_{\parallel}$ = 0.35 Hz V$^{-1}$ cm for the EF components, respectively [39].

The observation of two resonances in the recorded ODMR spectra is consistent with the geometry of the device, since the EF within the active region is mainly parallel to the axis joining the graphitic electrodes, i.e. the <100> crystal direction (**Fig. 2a**). Therefore, the four possible orientations of the NV defects are equivalent with respect to the EF direction, and thus determine a 4-fold degeneracy in the respective spin resonances, as well as the vanishing of the term $\delta$ in Eq. (1). Furthermore, the Stark-shifting effect of the parallel component of the EF was also disregarded since $d_{\parallel} \ll d_{\perp}$, leading to a negligible value of the term $\beta$. In order to exclusively account for the EF contribution to the recorded ODMR spectra, the splitting value measured at zero bias after the first voltage sweep (i.e. ~2.36 MHz, likely due to both the environmental magnetic field and local mechanical stresses surrounding the graphitized regions [40]) was subtracted from all the experimental values. This operation is equivalent to setting $\alpha$ = 0 in Eq. (1), as well as to disregarding the static strain *F*.

$E_x$ can therefore be expressed as a function of the ODMR splitting $\Delta w$ as $E_x = \Delta w / (2 d_{\perp})$ [32]. Considering that the EF forms a $\pi/4$ angle with the NV axes, its total intensity is then $E = \Delta w/(2\sqrt{2} d_{\perp})$. It is worth noting that Joule heating effects are included in the temperature-dependent $H_0$ term in Eq. (1) and result in the same shift for both the $m_s = +1$ and $m_s = -1$ resonances towards lower frequencies [27]. As the EF is evaluated as the difference between the $m_s = 0$ and $m_s = \pm 1$ transition resonances, its estimation is therefore insensitive to temperature variations. Furthermore, considering a temperature dependence given by $H_0 = D + b_T \Delta T$, where $D$ = 2.87 GHz at room temperature and $b_T$ = 74.2 kHz K$^{-1}$ [27,39], no trends in the ODMR resonances shift associated with a temperature increase were observed within the experimental sensitivity (~5 MHz FWHM of the ODMR resonances), posing n upper limit for the variation in the internal temperature to $\Delta T \cong 70$ K.

The estimated EF norm at the probed point is reported in **Fig. 2d** as a function of the applied bias voltage. The blue line represents the result of a finite element method (FEM) simulation performed assuming a relative dielectric constant $\varepsilon_r$ = 7.6, as estimated in a previous electrical characterization of the same device [33].

As expected, for applied bias voltages below $V_T$ (i.e. in the ohmic conduction regime) the EF linearly increases with the applied bias voltage at both polarities. However, similarly to what observed in the current-voltage curves (see **Fig. 1c**), its trend is not symmetric under polarity reversal. The EF estimation at positive bias voltages (i.e. first voltage sweep) significantly differs from the theoretical expectations, indicating that the preliminary voltage sweeps carried on the sample before this measurement produced a space charge build-up near the cathode, resulting in a ~40% increase of the internal EF under positive polarity. On the other hand, the good agreement between EF estimation and simulations under negative applied bias (i.e. second voltage sweep) indicate that the previous operating of the device in the PF conduction regime ($V_{bias} > +V_C$) provided a "reset" of the electron traps in the active region (the "reset" state of the device at zero bias after the first voltage sweep is labeled as "[R]" between **Figs. 2b-2d**).

After a progressive deviation from the linear behavior observed in the +125-245 V range, the estimated local EF reaches a maximum value at a positive bias voltage of $V_T \sim$ 245 V. At higher biases, the EF steadily decreases up to $V \sim V_C$, above which the data suggest a new increase (**Figs. 2b-d**). Such a non-monotonic trend provides a direct insight into the conduction mechanism of the junction [37]. Remarkably, for bias voltages larger than ∼250 V the local EF is lower than what predicted by FEM simulations: this evidence indicates that the device does not experience a progressive detrapping of the above-mentioned electrons trapped near the cathode, but rather the build-up of a counter-field, associated with the widening of the negative space charge towards the anode. The decrease in the internal electric field is also in agreement with the macroscopic observation of the pronounced current increase (**Fig. 1c**), which is strictly connected to the high recombination current causing the electroluminescent emission at $V>V_C$.

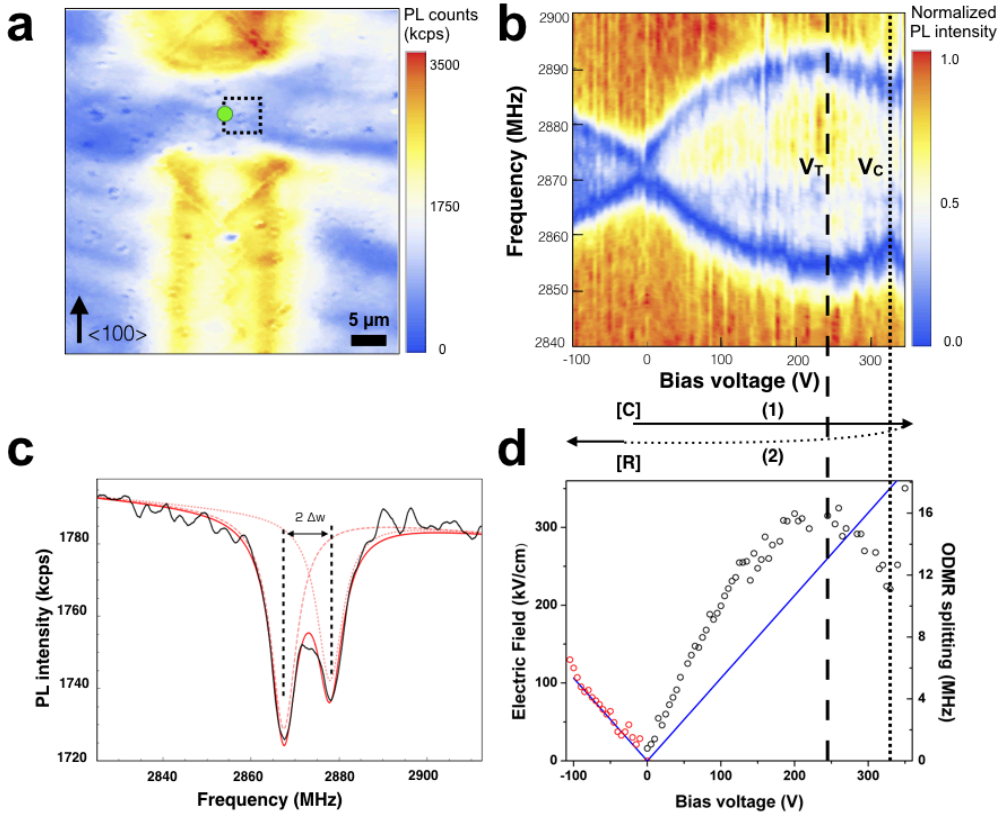

**Figure 2**: **a)** PL map of the unbiased device. **b)** Normalized ODMR spectra acquired during a voltage sweep from the position highlighted by the green spot in Fig. 2a. The black lines indicate the bias voltage $V_T$ (dashed) at which the ODMR splitting starts decreasing, as well as the bias voltage $V_C$ (dotted) corresponding to the transition to the high-current regime. **c)** Typical ODMR spectrum acquired at +25 V bias voltage (black line). The fitting curve (solid red) is superimposed, together with the two corresponding Lorentzian peaks. **d)** EF norm vs applied bias voltage, as evaluated from the ODMR splitting. The blue line represents the EF resulting from FEM simulations.

*EF mapping*. In order to evaluate the spatial distribution of the EF in within the active region, maps of the ODMR signals were acquired by raster-scanning the sample position over the 5×5 µm² area highlighted by the black square in **Fig. 2a**. The measurements were carried after the second voltage sweep, i.e. after having polarized the device at $V_{bias}$ = 105 V. To investigate possible polarization effects, a map at $V_{bias}$ = -25 V was followed by a second one at $V_{bias}$ = 0 V. **Figs. 3a-b** display the maps of the EF norm evaluated from the ODMR resonances, as previously described. A 2D FEM simulation based on the solution of the Poisson's equation, under the simplified assumption of a homogeneous space charge concentration of $-2.5 \times 10^{-5}$ C cm⁻³ (estimated from a linear interpolation of the filled trap density ρ, assuming the complete filling, i.e. ρ = $2 \times 10^{15}$ cm⁻³ occurs at $V=V_C$ [21]) was carried for $V_{bias}$ = -25 V to provide a qualitative comparison with the measured EF distribution (**Fig. 3c**, in which the black square highlights the mapped region). The features of the experimental and FEM-simulated EF maps are in satisfactory overall qualitative agreement, while at the same type displaying substantial differences, thus indicating the build-up of spatial patterns of charge density in the active region. The average EF intensity measured from the ODMR maps is significantly higher (median: 79.4 kV cm⁻¹) with respect to both the FEM simulation and the estimation obtained during the previous voltage sweep (~40 kV cm⁻¹, **Fig. 2d**). This evidence highlights the role of electron trapping in the space charge build-up caused by large negative voltages ($V_{bias}$ = -100 V, see **Fig. 2d**), after having "erased" the device at large positive bias voltages. This interpretation is further strengthened by the significant residual EF (median map value: 73.7 kV cm⁻¹) observed at $V_{bias}$ = 0 V (**Fig. 3b**).

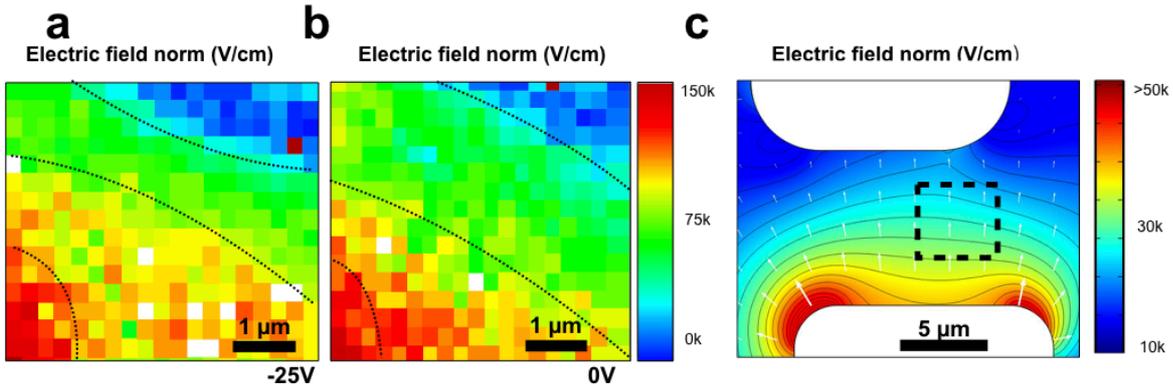

**Figure 3**: **a)**, **b)** EF maps from the 5×5 μm$^2$ region highlighted in Fig. 2a, as obtained from the ODMR spectra; black lines represent the isofield lines. **c)** 2D FEM simulation of the EF norm distribution at $V_{bias}$ = -25 V. The black square highlights the mapped region. White arrows represent the EF vector, black lines the isofield lines. The white regions represent the electrodes.

## IV. CONCLUSIONS

We demonstrated that the ODMR readout from ensembles of NV centers allows the investigation of the local effects of charged traps on the internal EF of a diamond-graphite-diamond junction. This analysis highlighted the formation of a space charge distribution and resulted in the direct observation of memory effects associated with the sample biasing history. This direct experimental insight represents a significant progress with respect to previous works, in which complex charge transport dynamics in defective diamond were indirectly inferred on the basis of electrical and optical characterizations [19-21].

In particular, the imaging of the internal EF via ODMR highlights the strong persistence of long-lived charged traps in ion-irradiated diamond, as well as providing a new insight in their role in the polarization effects that were widely reported (at a macroscopic scale) in the literature by means of conventional electrical characterization.

In perspective, this approach offers a new tool to investigate the electronic features and conduction mechanisms in different classes of diamond devices, ranging from radiation detection [41-42] to bistable systems [43,44]. Furthermore, it provides a complementary technique to assess the radiation hardness of diamond-based detectors, i.e. the progressive modification of the internal electric field in their active regions as a function of the lattice damage [7,45,46, 47].
The future exploitation of advanced ODMR sensing protocols [10,30,47-49] combined with the use of advanced fabrication techniques for the creation of graphitic nanochannels [50] opens interesting perspectives for the investigation of the effects of individual defects [32,51], with a potential impact in the development of integrated devices for quantum technologies.

## ACKNOWLEDGEMENTS

This research was supported by the following projects: "DIESIS" project funded by the Italian National Institute of Nuclear Physics (INFN) - CSN5 within the "Young research grant" scheme; Coordinated Research Project "F11020" of the International Atomic Energy Agency (IAEA).